\documentclass[conference]{IEEEtran}
\IEEEoverridecommandlockouts
\usepackage{cite}
\usepackage{amsmath,amssymb,amsfonts}
\usepackage{algorithmic}
\usepackage{graphicx}
\usepackage{textcomp}
\usepackage{xcolor}
\def\BibTeX{{\rm B\kern-.05em{\sc i\kern-.025em b}\kern-.08em
    T\kern-.1667em\lower.7ex\hbox{E}\kern-.125emX}}
\graphicspath{ {Figures/} }
\begin{document}

\title{Proof of Reference(PoR): A unified informetrics based consensus mechanism\\}

\author{\IEEEauthorblockN{1\textsuperscript{st} Parul Khurana}
\IEEEauthorblockA{\textit{School of Computer Applications} \\
\textit{Lovely Professional University}\\
Phagwara, India \\
parul.khurana@lpu.co.in}
\and
\IEEEauthorblockN{2\textsuperscript{nd} Geetha Ganesan}
\IEEEauthorblockA{\textit{Advanced Computing Research Society} \\
\textit{Porur}\\
Chennai, India \\
gitaskumar@yahoo.com}
\and
\IEEEauthorblockN{3\textsuperscript{rd} Gulshan Kumar}
\IEEEauthorblockA{\textit{School of Computer Science and Engineering} \\
\textit{Lovely Professional University}\\
Phagwara, India \\
gulshan.16865@lpu.co.in}
\and
\IEEEauthorblockN{4\textsuperscript{th} Kiran Sharma}
\IEEEauthorblockA{\textit{School of Engineering and Technology} \\
\textit{BML Munjal University}\\
Gurugram, India \\
kiran.sharma@bmu.edu.in}
}

\maketitle

\begin{abstract}
Bibliometrics is useful to analyze the research impact for measuring the research quality. Different bibliographic databases like Scopus, Web of Science, Google Scholar etc. are accessed for evaluating the trend of publications and citations from time to time. Some of these databases are free and some are subscription based. Its always debatable that which bibliographic database is better and in what terms. To provide an optimal solution to availability of multiple bibliographic databases, we have implemented a single authentic database named as ``conflate'' which can be used for fetching publication and citation trend of an author. To further strengthen the generated database and to provide the transparent system to the stakeholders, a consensus mechanism ``proof of reference (PoR)'' is proposed. Due to three consent based checks implemented in PoR, we feel that it could be considered as a authentic and honest citation data source for the calculation of unified informetrics for an author.

\end{abstract}


\section{Introduction}\label{AA}
Funding pressures, promotion decisions, research reputations, and global competitiveness have always reminded the research community to use bibliometrics for evaluations. Such metrics has become an open topic of debate for sharing of expertise with high level analysis \cite{pringle2008trends}. In short span of time, bibliometrics is emerged as the useful tool to analyse the influence of research outputs, complementing qualitative research indicators and assisting in the evaluation of research quality and impact \cite{umate41bibliometric}. With the availability of major bibliographic databases like Scopus, and Web of Science, it has become a trend now to compile and produce data for the statistics based on bibliometrics \cite{archambault2009comparing}.

The comparison of Scopus and Web of Science is been the focus of number of authors. Both bibliographic databases are compared on the basis of their content comprehensiveness \cite{martin2018google} and different perspective of calculating citations based on different categories \cite{khurana2021impact}. Being traditional databases, both Scopus and Web of Science are also compared with Google Scholar, Microsoft Academic\cite{visser2021large}, OpenCitations \cite{martin2021google}, Crossref and Dimensions \cite{harzing2019two} by authors. Despite of multiple bibliographic databases it is always in debate that which bibliographic database is better and in what terms \cite{martin2019google}.

During the time of interviews, recruiting agencies often asks the authors to enter Scopus and Web of Science credentials separately, as a matter of fact there is no common platform which can present single $h$-index, publication, and citation count based on multiple bibliographic databases. A novel system presenting unified informetrics of an author named as ``Conflate'' \cite{khurana2021weighted} is presented. This system in based on the publication and citation data extracted from Scopus and Web of Science which performs citation analysis and presents single authentic $h$-index, publication, and citation count of an author across multiple bibliographic databases. System is developed in such a way that it can be used with $n$ number of bibliographic databases as well.

To strengthen the authentic system of ``conflate'', it is empowered with the concept of distributed ledger technology (DLT). DLT has emerged as a solution based technology which can overcome the limitations faced in various fields. The most common fields where DLT may be seen are accessing of services \cite{singh2020novel}, verification of academic records \cite{aamir2020blockchain}, validation and authorization of certificates \cite{kinkelin2020hardening}, sharing of credentials \cite{mishra2020implementation}, and in mobile higher education \cite{arndt2020blockchain}. Such implementations give us a thought and an idea of proposing DLT in publication industry. This is novel approach and can provide a robust, highly transparent, and decentralized platform for the presentation of unified informetrics of an author to its stakeholders.

The objectives of our study are:
\begin{itemize}
\item To identify the strengths of distributed ledger technology.
\item To propose a consensus mechanism named as ``proof of reference''.
\end{itemize} 

The study is organized as follows: Section~\ref{MR} is on methodology and results based  on citation analysis, proof of reference and implementation. Summary is presented in Section~\ref{SU}
%~~~~~~~~~~~~~~~~~~~~~~~~~~~~~~~~~~~~~~~~~~~~~~~~~~~~~~~~~~~~~~~~~~~~~~~~~~~~~~~~~~~~~~~~~

%~~~~~~~~~~~~~~~~~~~~~~~~~~~~~~~~~~~~~~~~~~~~~~~~~~~~~~~~~~~~~~~~~~~~~~~~~~~~~~~~~~~~~~~~~
\section{Methodology and Results}
\label{MR}
%~~~~~~~~~~~~~~~~~~~~~~~~~~~~~~~~~~~~~~~~~~~~~~~~~~~~~~~~~~~~~~~~~~~~~~~~~~~~~~~~~~~~~~~~~
\subsection{Citation Analysis}\label{CA}
\begin{figure}[htbp]
\centerline{\includegraphics[width=0.85\linewidth]{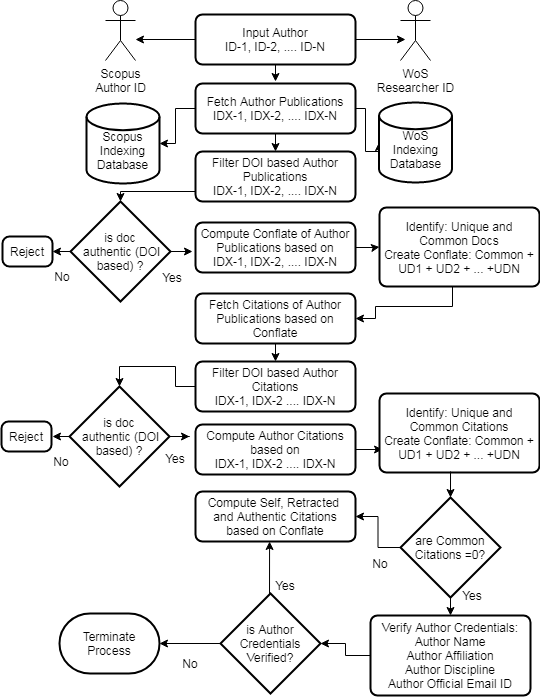}}
\caption{Flowchart demonstrates the process of generating conflate of an author based on data extracted from Scopus and WoS.}
\label{fig1}
\end{figure}

%~~~~~~~~~~~~~~~~~~~~~~~~~~~~~~~~~~~~~~~~~~~~~~~~~~~~~~~~~~~~~~~~~~~~~~~~~~~~~~~~~~~~~~~~~
Bibliographic databases like Web of Science, and Scopus are frequently used to procure citation counts. These are paid databases and often require a subscription for the extraction of citation data. Although there are free bibliographic databases exist, like Google Scholar etc., these two have proved their worth as an authentic source of bibliographic data. Both have their unique way of recording and counting citations with their own scope of content coverage as well \cite{kulkarni2009comparisons, adriaanse2013web}. Hence, by considering both Web of Science and Scopus as an authentic source of information, a solution named as ``Conflate'' has been presented as a unified informetrics \cite{khurana2021weighted}. Fig.~\ref{fig1} shows the calculation steps of ``Conflate'' for an author.
In the first step, an author is required to enter his credentials such as AuthorID in case of Scopus and ResearcherID in case of Web of Science. After verifying the input received from an author, system will extract the author publications from both bibliographic databases. To filter the authentic author publications, a system will select the publications with DOI and reject the publications with no DOI information associated with them. DOI is a permanent identifier used to uniquely identify a publication, text etc. in online platform which helps the publication in its operability and interoperability. \cite{gorraiz2016availability}. A publication is considered as an authentic publication if it carries valid DOI, hence helping in the production of authentic database of ``Conflate''. After filtering the unique and common publications in Scopus and Web of Science, conflate of author publications will be computed presenting the unified publication count among both bibliographic databases.

In the next major step, a unified publication count will be used to extract citation data from both bibliographic databases. After extracting the citation data, a second level of DOI based filtration will be applied. A system will keep the citations with valid DOIs and reject the citations with no DOI information associated with them. System will also identify the self author citations, and retracted citations for the final computation of citation count of an author. Hence, helping in the production of the system with authentic and true citation count for an author. In the next step, system will identify the unique and common citations extracted from the database, thus generating the audit of citations in case of any major observations like common citation count = 0. Such cases are treated with the further audit of author credentials at organization	level as well.

%~~~~~~~~~~~~~~~~~~~~~~~~~~~~~~~~~~~~~~~~~~~~~~~~~~~~~~~~~~~~~~~~~~~~~~~~~~~~~~~~~~~~~~~~~
\subsection{Proof of Reference (PoR)}\label{POR}
%~~~~~~~~~~~~~~~~~~~~~~~~~~~~~~~~~~~~~~~~~~~~~~~~~~~~~~~~~~~~~~~~~~~~~~~~~~~~~~~~~~~~~~~~~
Ledgers have been at the heart of trade since the dawn of time, and they are used to record a wide range of transactions, the majority of which include the storage of personal property and currency. They were encouraged to write on clay tablets, which were then transferred to paper, vellum, and papyrus \cite{chatzopoulos2021towards}. However, computerization, which began as a conversion from manuscript to bytes, is the only notable innovation. Systems aid in the construction of digital ledgers by providing features and capabilities that go beyond those of traditional manual ledgers \cite{puntinx2017distributed}.

A distributed ledger is a data bank that may be accessed from multiple layouts or groups in a system. Each member of a linkage may have their own personal copy of the ledger. Any change to the ledger is instantly reflected throughout the system \cite{perdana2021distributed}. It doesn't matter if the entities are monetary, legal, or physical.

Typical DLT applications may include, providing benefits such as passports and documenting property deeds, as well as improving government services, improving health care by enhancing services and ensuring the safe transfer of medical histories and records, refining facilities that place a high value on transactions that are sluggish, exclusive, and reliant on intermediaries, use of smart contracts in many divisions, including as nutrition, liveliness, pharmaceuticals, aeronautics, communications, and transportation for expansion, assisting in the facilitation of equitable compensation in the fashion industry, improving the IoT business and expertise replication, improving personality management by ensuring that identities are secure and transferable, applications in economic services, company structure efficiency, and risk management, assuring the integrity of composite organization arrangements and mechanisms. Thus DLT has the potential to transform the supply of social amenities with increasing productivity. It's a database that can securely store both physical and financial assets for speedy network sharing \cite{brown2016distributed}.

\begin{figure}[htbp]
\centerline{\includegraphics[width=0.85\linewidth]{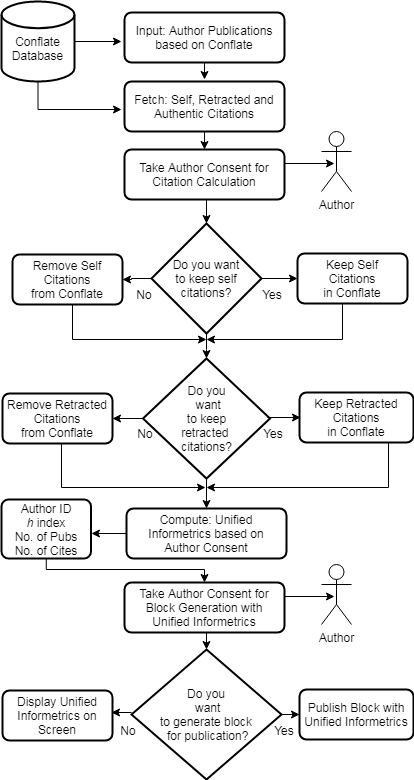}}
\caption{Flowchart demonstrates the process of consensus mechanism - ``proof of reference'' designed to strengthen the conflate database of an author.}
\label{fig2}
\end{figure}

DLT is built on consensus protocols. These protocols decide the new possibilities of well optimized solutions given by this technology. Consensus means agreement which all stakeholders must accept while taking a decision wherever necessary. Traditional consensus protocols include proof of work, proof of stake, and proof of authority \cite{sankar2017survey} etc. Different authors have also presented the typical consensus protocols like proof of trust negotiation \cite{feng2019potn}, proof of honesty \cite{makhdoom2020pledge}, proof of useful work \cite{baldominos2019coin}, proof of credit \cite{han2019fair}, proof of reputation \cite{zhuang2019proof}, proof of benefit \cite{liu2019proof}, proof of block and trade \cite{biswas2019pobt}, proof of disease \cite{talukder2018proof}, proof of download \cite{costa2020blockchain}, proof of federated learning \cite{qu2021proof}, proof of game \cite{kumar2021proof}, proof of kernel work \cite{lundbaek2018proof}, proof of learning \cite{bravo2019proof}, proof of luck \cite{milutinovic2016proof}, proof of play \cite{yuen2019proof}, proof of previous transactions \cite{xiang2019proof}, proof of QoS \cite{yu2019proof}, proof of reputation \cite{gai2018proof}, proof of search \cite{shibata2019proof}, proof of vote \cite{li2017proof}, proof of witness presence \cite{pournaras2020proof}, and proof of x-repute \cite{wang2020proof} etc. Summary of consensus mechanisms is given in Table~\ref{Table:1}.
%%%%%%%%%%%%=== Table 1 ===%%%%%%%%%%%%%%%%%%%%%
\begin{table*}[t]
\centering
\caption{Summary of consensus mechanisms}
\begin{tabular}{|c|c|l|}
\hline
\textbf{Year} & \textbf{Consensus Mechanism}            & \textbf{Index Terms Used}                                                                                                                                                                                                          \\ \hline
2016          & Proof of Luck                           & Blockchain, Trusted Execution Environments, Consensus Protocol, Intel Sgx                                                                                                                                                          \\ \hline
2017          & Proof of Vote                           & Blockchain, Consortium Blockchain, Consensus, Voting Mechanism                                                                                                                                                                     \\ \hline
2018          & Proof of Disease                        & \begin{tabular}[c]{@{}l@{}}Blockchain, Healthcare, Chatbot, Iot, Genomics, Precision Medicine, P6 Medicine,\\ Proof Of  Disease\end{tabular}                                                                                       \\ \hline
2019          & Proof of Benefit (PoB)                  & \begin{tabular}[c]{@{}l@{}}Eectric Vehicles, Distributed Trading System, Transactional Energy, Blockchain Technology,\\ Consensus Mechanism, Smart Grid\end{tabular}                                                               \\ \hline
2019          & Proof of Block \& Trade (PoBT)          & \begin{tabular}[c]{@{}l@{}}Internet Of Things, Blockchain, Scalability, Consensus, Transaction Rate, Ledger Size,\\ Distributed Ledger Technology, Interoperability\end{tabular}                                                   \\ \hline
2019          & Proof of Credit                         & \begin{tabular}[c]{@{}l@{}}Blockchain, Consensus Protocol, Hybrid Incentive Mechanism, Proof Of Credit (Poc),\\ Proof Of Stake (Pos)\end{tabular}                                                                                  \\ \hline
2019          & Proof of Federated Learning             & Blockchain, Consensus Mechanism, Federated Learning, Reverse Game                                                                                                                                                                  \\ \hline
2019          & Proof of Learning                       & Distributed Consensus, Blockchain, Machine Learning, Incentive Mechanisms, Algorand                                                                                                                                                \\ \hline
2019          & Proof of Play                           & Blockchain, Security, Consensus Model, P2P, Games                                                                                                                                                                                  \\ \hline
2019          & Proof of Previous Transactions (PoPT)   & Blockchain, Consensus Algorithm, Jointcloud, Parallel Accounting                                                                                                                                                                   \\ \hline
2019          & Proof of QoS                            & Blockchain, Consensus Protocol, Bft, Quality Of Service, Proof-Of-QoS (PoQ)                                                                                                                                                        \\ \hline
2019          & Proof of Reputation                     & \begin{tabular}[c]{@{}l@{}}Blockchain, Bitcoin, Ethereum, Consensus, Decentralization, Ncdawarerank, Hodgerank,\\ Pagerank, Proof Of Work, Proof Of Stake, Delegated Proof Of Stake, Delegated Proof Of \\ Reputation\end{tabular} \\ \hline
2019          & Proof of Search                         & Peer-To-Peer Computing, Distributed Computing, Grid Computing                                                                                                                                                                      \\ \hline
2019          & Proof of Trust Negotiation              & Blockchain, Consensus Protocol, Trust Management, Miners Selection, Iot                                                                                                                                                            \\ \hline
2019          & Proof of Useful Work                    & \begin{tabular}[c]{@{}l@{}}Blockchain, Smart Contract, Crowd Computing, Decentralised System, Large Scale\\ Computing\end{tabular}                                                                                                 \\ \hline
2020          & Proof of Assets and Proof of Reputation & \begin{tabular}[c]{@{}l@{}}Consortium Blockchain, Consensus Algorithm, Voting, Deposit, Reputation, Verifiable\\ Random Function\end{tabular}                                                                                      \\ \hline
2020          & Proof of Download                       & Blockchain, Repository, Distributed Consensus                                                                                                                                                                                      \\ \hline
2020          & Proof of Honesty                        & \begin{tabular}[c]{@{}l@{}}Blockchain Consensus, Byzantine Fault Tolerance, Distributed Consensus, Proof Of  Honesty,\\ Miner Selection\end{tabular}                                                                               \\ \hline
2020          & Proof of Witness Presence               & Augmented Democracy, Blockchain, Consensus Mechanism, Smart City, Witness Presence                                                                                                                                                 \\ \hline
2020          & Proof of X Repute                       & Blockchain, Consensus Protocols, Iot, Trust Management, Repute Incentive                                                                                                                                                           \\ \hline
2021          & Proof of Game (PoG)                     & \begin{tabular}[c]{@{}l@{}}Blockchain, Consensus Algorithm, Proof Of Concepts, Trusted Environments, Cryptography, \\ Cryptocurrency\end{tabular}                                                                                  \\ \hline
2021          & Proof of Kernel Work                    & Cryptographic Sortition, Access Control, Blockchain, Cybersecurity                                                                                                                                                                 \\ \hline
\end{tabular}
\label{Table:1}
\end{table*}
%%%%%%%%%%%%%%%%%%%%%%%%%%%%%%%%%%%%%%%%%%-------------------------------------------------
To enhance the performance of citation analysis, a ``proof of reference'' based consensus mechanism is proposed which is based on conflate database generated for an author, see Fig.~\ref{fig2} for the detailed step wise description of mechanism.
Conflate database generated in Fig.~\ref{fig1} is used as an input for the generation of blocks in DLT. Initially in the implementation of ``proof of reference'' mechanism, author publications are given as an input to the system with citations data, broadly divided into three categories, self citations, retracted citations and authentic citations. For the calculation of actual citations of an author, a consent of author will be taken to consider self citations as an actual citations, if an author agrees with the consent, self citations will be considered otherwise self citations will be discarded. Self citations are the citations where an author may cite his own piece of work for rewards, funding, and better position etc. \cite{da2021right}. In next step, author will be asked to consider retracted citations, these retracted citations may include falsification of the data, ethical and logical misconduct as well. It is the responsibility of an author here to select the citations which are completely genuine and authentic piece of work for his final count of citations \cite{sharma2020patterns}. If an author agrees, retracted citations will be considered otherwise these will be rejected.

Based on author inputs received for consideration of self and retracted citations, a unified author metrics will be calculated presenting author $h$-index, author publications and author citations. In the last step, a final author consent will be taken for the generation of block for the publication in distributed ledger. If author agrees, block carrying unified informetrics of an author will be published otherwise an output will be displayed to an author on his screen. 

%~~~~~~~~~~~~~~~~~~~~~~~~~~~~~~~~~~~~~~~~~~~~~~~~~~~~~~~~~~~~~~~~~~~~~~~~~~~~~~~~~~~~~~~~~
\subsection{Implementation}\label{IM}
\begin{figure}[htbp]
\centerline{\includegraphics[width=0.85\linewidth]{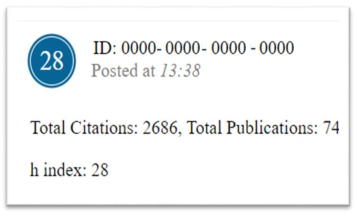}}
\caption{Example of an author block.}
\label{fig3}
\end{figure}

%~~~~~~~~~~~~~~~~~~~~~~~~~~~~~~~~~~~~~~~~~~~~~~~~~~~~~~~~~~~~~~~~~~~~~~~~~~~~~~~~~~~~~~~~~
Implementation of ``conflate'' has been done in two steps, in first step, data was extracted from Scopus \cite{pybliometrics} and Web of Science \cite{wos} and in second step, block generation was done by deploying DLT server on the local system \cite{ibm_blockchain_post}. Sample block generation of an author will look like Fig.~\ref{fig3}.
A Flask based framework is used to run decentralized network on multiple machines. Initially a DLT based server is setup on the system to run the scripts of deployment. A simple user interface based on Python, HTML, CSS and JavaScript was designed for receiving the inputs from an author. An application is written to allow an author to share his bibliometrics in an append mode, such that immutability and durability of presented data is achieved. The structure of data will include unified informetrics in the form of $h$-index, number of publications and number of citations. A mandatory time stamp will be affixed to uniquely identify the block request received from an author.

Implementation steps will include:
\begin{itemize}
\item grouping of transactions together into blocks.
\item stamping of the blocks with digital fingerprints.
\item building a chain of the blocks.
\item adding blocks to the chain using a Proof of Reference method.
\item designing of user interfaces.
\item establishing decentralization and consensus.
\item construction of application, and initialization of the program.
\end{itemize}

First time execution will include:
\begin{itemize}
\item cloning of project folder in local system.
\item installation of dependencies.
\item initialization of DLT node server at port 8080.
\item initialization of DLT based application on local system at port 5000.
\end{itemize}

%~~~~~~~~~~~~~~~~~~~~~~~~~~~~~~~~~~~~~~~~~~~~~~~~~~~~~~~~~~~~~~~~~~~~~~~~~~~~~~~~~~~~~~~~~
\section{Conclusions and Future Work}\label{SU}
%~~~~~~~~~~~~~~~~~~~~~~~~~~~~~~~~~~~~~~~~~~~~~~~~~~~~~~~~~~~~~~~~~~~~~~~~~~~~~~~~~~~~~~~~~
Proof of Reference is currently designed for strengthening the concept of citation analysis with the implementation support of DLT. Citation analysis is performed on the data extracted from Scopus and Web of Science at author level. Generation of authentic database for an author has been done based on publications and citations carrying valid DOIs only. Further PoR is divided into three authenticity checks which restricts the consideration of plagiarized work, forged authorship, and fabricated data for the calculation of unified informetrics of an author. The implemented system gives an opportunity to an author to develop a working prototype for authors, coauthors, reviewers and accreditation and ranking agencies to provide a unified authentic informetrics for promotion, funding and recognition of research initiatives. In the future, we will try to adapt PoR in possible areas of publication industry where there is a high demand for the authentic research credentials of an organizations as well.
\section{Conflict of interest}
The authors declare that they have no conflict of interest.
%~~~~~~~~~~~~~~~~~~~~~~~~~~~~~~~~~~~~~~~~~~~~~~~~~~~~~~~~~~~~~~~~~~~~~~~~~~~~~~~~~~~~~~~~~
\bibliographystyle{IEEEtran}
\bibliography{cas-refs}
\vspace{12pt}
\end{document}